\documentclass[reprint,amsmath,amssymb,aps,prb]{revtex4-2}

\usepackage{xcolor}

\usepackage{tabularx}
\newcolumntype{Y}{>{\centering\arraybackslash}X}

\usepackage[newcommands]{ragged2e}
\usepackage{svg}
\usepackage{graphicx}
\usepackage{dcolumn}
\usepackage{bm}
\usepackage{chemformula} % Formula subscripts using \ch{}
\usepackage[T1]{fontenc} % Use modern font encodings
\usepackage{comment}
\usepackage{physics}
\usepackage[utf8]{inputenc}
\usepackage[colorlinks=true,linkcolor=blue,filecolor=magenta,urlcolor=blue,citecolor=blue,pdfstartview=FitH]{hyperref}
\usepackage[justification=raggedright]{caption}
\usepackage{subcaption}
\usepackage{float}
\graphicspath{{images/}}
\usepackage{blindtext}
\usepackage{array}
\newcolumntype{P}[1]{>{\centering\arraybackslash}p{#1}}
\usepackage{subfiles}
\usepackage{amsmath}
\usepackage{amsfonts}
\usepackage{amssymb}

\usepackage{multirow}

\definecolor{clr1}{rgb}{0,0,0}
\definecolor{clr2}{rgb}{0.0,0.35,0.0}
\newcommand{\figref}[1]{Fig.~\ref{#1}}
\newcommand{\secref}[1]{SEC.~\ref{#1}}
\newcommand{\eqaref}[1]{Eq.~\eqref{#1}}
\newcommand{\apref}[1]{Appendix~\ref{#1}}
\newcommand{\at}[2][]{#1|_{#2}}
\newcommand{\degr}{^{\circ}}
\newcommand{\tref}[1]{TABLE~\ref{#1}}
\newcommand{\beq}{\begin{equation}}
	\newcommand{\eeq}{\end{equation}}
\newcommand{\ie}{i.e.,}

\newcommand{\bse}{\begin{subequations}}
	\newcommand{\ese}{\end{subequations}}
\newcommand{\bea}{\begin{eqnarray}}
	\newcommand{\eea}{\end{eqnarray}}
\newcommand{\bem}{\begin{displaymath}}
	\newcommand{\eem}{\end{displaymath}}
\newcommand{\bmat}{\begin{bmatrix}}
	\newcommand{\ebmat}{\end{bmatrix}}

\newcommand{\bc}{\begin{center}}
	\newcommand{\ec}{\end{center}}

\newcommand{\squaret}[1]{\sqrt{#1}}

\begin{document}
	\title{Electronic analogue of Fourier optics with massless Dirac fermions scattered by quantum dot lattice}
%	\title{Fourier Electron Optics with Massless Dirac Fermions Scattered by Quantum Dot Lattice}
	
	\author{Partha Sarathi Banerjee}
	\affiliation{Department of Physics,
		Indian Institute of Technology Delhi,
		Hauz Khas, New Delhi 110016}

	\author{Rahul Marathe}
	\affiliation{Department of Physics,
		Indian Institute of Technology Delhi,
		Hauz Khas, New Delhi 110016}
	
	\author{Sankalpa Ghosh}
	\thanks{Corresponding author. Email: \color{blue}sankalpa@physics.iitd.ac.in}
%	\email{Corresponding author.	Email: sankalpa@physics.iitd.ac.in}
	\affiliation{Department of Physics,
		Indian Institute of Technology Delhi,
		Hauz Khas, New Delhi 110016}
	
%	\date{February 17, 2024}

	\begin{abstract}
		The field of electron optics exploits the analogy between the movement of electrons or charged quasiparticles, primarily in two-dimensional materials subjected to electric and magnetic (EM) fields and the propagation of electromagnetic waves in a dielectric medium with varied refractive index. We significantly extend this analogy by introducing an electronic analogue of Fourier optics dubbed as Fourier electron optics (FEO) with massless Dirac fermions (MDF), namely the charge carriers of single-layer graphene under ambient conditions, by considering their scattering from a two-dimensional quantum dot lattice (TDQDL) treated within Lippmann-Schwinger formalism. By considering the scattering of MDF from TDQDL with a defect region, as well as the moir\'{e} pattern of twisted TDQDLs, we establish an electronic analogue of Babinet's principle in optics. Exploiting the similarity of the resulting differential scattering cross-section with the Fraunhofer diffraction pattern, we construct a dictionary for such FEO. Subsequently, we evaluate the resistivity of such scattered MDF using the Boltzmann approach as a function of the angle made between the direction of propagation of these charge-carriers and the symmetry axis of the dot-lattice, and Fourier analyze them to show that the spatial frequency associated with the angle-resolved resistivity gets filtered according to the structural changes in the dot lattice, indicating wider applicability of FEO of MDF.
	\end{abstract}
	
	\maketitle
	
%	\section{Introduction}

	\section{Introduction}

	The unique transmission properties of massless Dirac fermions (MDF) in graphene through potential landscapes created by a variety of electromagnetic (EM) fields, particularly in the ballistic regime \cite{Cheianov2006, kat, Cheianov2007, young2009quantum, Stander2009, Chen2016}, and its similarity with the light transmission through an optical medium with unconventional dielectric properties such as metamaterials \cite{Pendry2000, Schurig2006} make graphene an excellent material to realize electron optics-based devices in a solid state system. The realisation of negative refraction \cite{lee2015observation}, chiral Veselago lensing of MDF in two \cite{brun2019} and three dimensions \cite{Tchoumakov2022}, tunable Veselago interference in a bipolar graphene microcavity \cite{zhang2022gate},  creation of a Dirac fermion microscope \cite{boggild2017}, collimation \cite{Park2008, Liu2017, wang2019graphene},  and different type of interferometers \cite{wei2017mach,jo2021quantum,rickhaus2013ballistic, Forghieri2022}, gate tunable beam-splitter of such MDF \cite{Rickhaus2015}, Fabry-p\'erot resonator in graphene/hBN moir\'e super-lattice \cite{handschin2017fabry}, gradient index electron optics in graphene p-n junction \cite{Paredes2021}, Mie scattering in graphene, \cite{Heinisch2013} are few milestones in this direction.  Most of these experimental and theoretical studies are based on theoretical modelling of Dirac fermions scattered by the potential, which are constant in one direction \cite{Park2008, Sajjad2012, Cheianov2006, Anwar2020, Cheianov2007,young2009quantum, Chen2016, kat, Stander2009, Ghosh2009, Sharma2011, Zhao2023, Karalic2020}, and hence limit the range of applications in this fast-growing field.

\begin{figure*}
		\centering
		\includegraphics[width=  \linewidth]{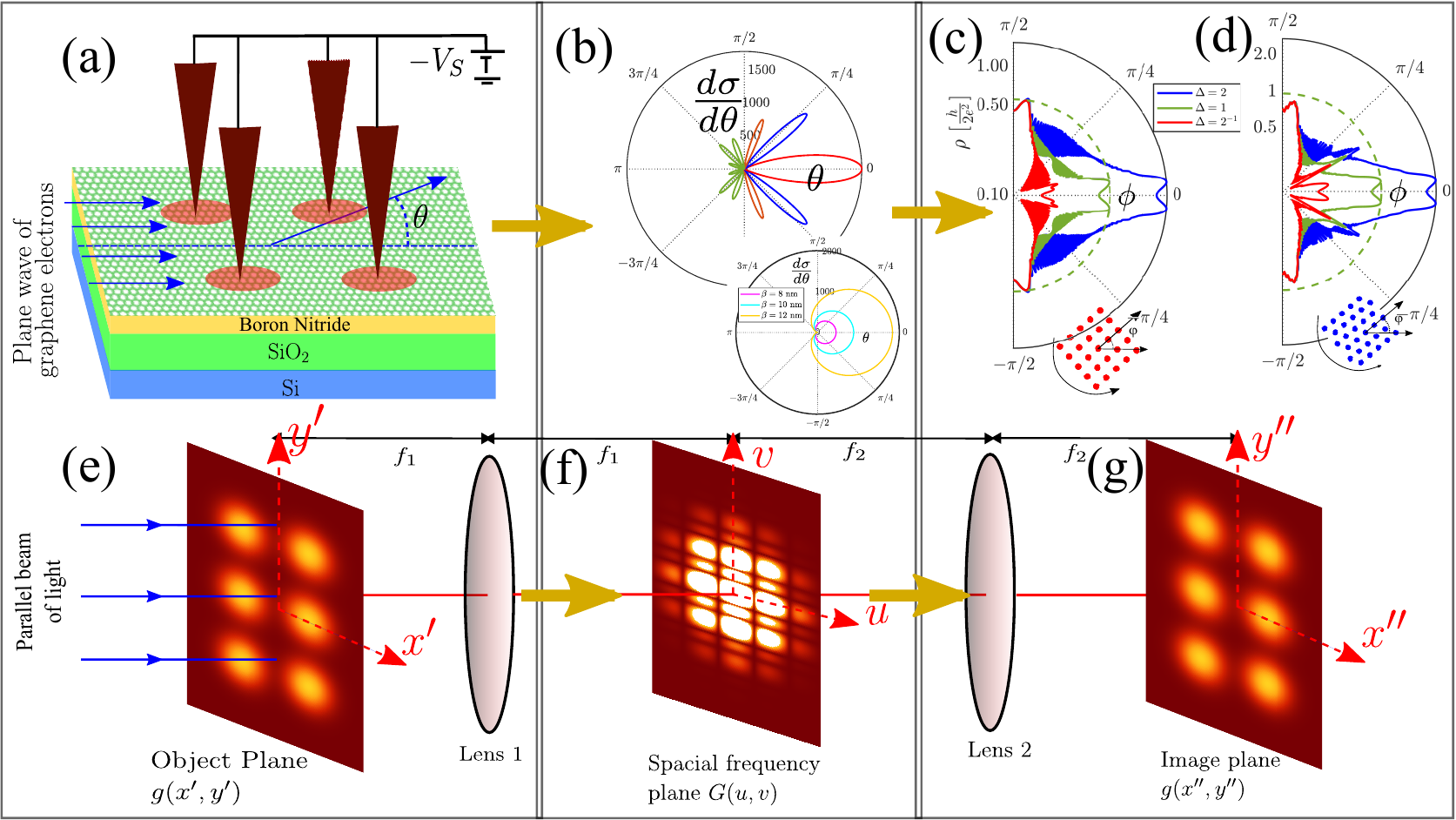}
		\caption{\justifying (a) The schematic diagram of a plane wave (direction shown in blue arrows) of charge carriers in ballistic graphene that are modelled as MDF under ambient conditions, getting scattered by a two-dimensional array of Gaussian quantum dot(QD) potentials created by STM tips. (b) The polar plot of the DSC for a square lattice of QDs as given by \eqaref{square}, of dimension $N_1=10$,$N_2=0$ and orientation $\phi=0$. The central maxima at $\theta=0$ is multiplied with $1.6\times 10^{-3}$ for better visibility. The first maxima on both sides are multiplied by $0.8\times 10^{-2}$. The second ones are multiplied by $0.32$ for better visibility with respect to the other smaller peaks. In the inset we have shown the differential scattering cross section for a single QD for differential values of $\beta$. 
		In (c) and (d) the angle-resolved dc-resistivity of the system parallel to the direction of propagation of the incoming plane wave of graphene electrons is plotted under this scattering potential rotated at an arbitrary angle.   The resistivity pattern for square and hexagonal lattices of QDs is shown in (c) and (d) for $N_2=100$. The resistivity at $\phi=0^{\circ}$ and $90^{\circ}$ is the same for the square lattice but not in the case of the hexagonal lattice. In Figs. (e)-(g) we compare the process described in (a)-(d) with the two-dimensional optical spatial frequency processor, whereas a short thesaurus listing various analogue quantities in these two systems is given in  \tref{table}.
		In (e) we show that the object is positioned in the front focal plane of lens 1. The Fourier transform of the object distribution is found in the back focal plane of lens 1 as shown in (f). This plane is called spatial frequency plane \cite{Birch1968, KGBirch_1972}. At the image plane in (g) the object distribution is recovered.
		 %Particularly in (e), we show the field distribution in the image plane. The Fraunhofer diffraction pattern of the image is shown in (f), which is the Fourier transform of the field distribution of the image plane forming spacial frequency plane \cite{Birch1968, KGBirch_1972} whereas in (g), the Fraunhofer diffraction pattern \ie the Fourier transform of the spatial frequency plane is shown.
	}
		\label{figSL}
	\end{figure*}

	New possibilities can emerge if the EM potential that can scatter such MDF can vary along both transverse directions. For example, in the well-known Fraunhofer diffraction, when the observation point is significantly distant($z$) from the diffracting object dubbed as the far-field case, the field distribution at the observation plane is the Fourier transform of the aperture function($A(x', y')$) \cite{duffieux1983fourier,Birch1968, ghatak2009optics, goodman2005introduction, KGBirch_1972, BornWolf, jenkins1976fundamentals}.
%	\color{blue}
	The diffracting object is positioned at the front focal plane of a lens, resulting in the generation of a Fourier transform of the object at the back focal plane of the lens,
%	\color{black}
	% The diffracting object is positioned at the back focal length of a lens, resulting in the generation of an image at the front focal length of the lens,
	thereby satisfying the conditions for the Fraunhofer approximation$\left(z >> \left[x'^2+y'^2\right]_{\text{max}}/\lambda \right)$ as depicted in \figref{figSL} and described in \tref{table}.
%	\color{blue}
	The framework of Fourier optics was introduced in a seminal work  by P. M. Duffieux \cite{duffieux1983fourier} and later elucidated in detail in the classic textbook of Born and Wolf \cite{BornWolf}.
%	\color{black}
	In this work we show that an electronic analogue of this situation can be realized in a fully two-dimensional (2D) scattering model, where the scattering of such MDF takes place from a two-dimensional superlattice potential that can be realised by creating an electrostatically defined array of quantum dots (QDs) on the surface of single-layer graphene \cite{silvestrov2007quantum, joung2011coulomb, on-off, 
		weddingcake, grushevskaya2021electronic, lee2016imaging, Zhao, li2022recent,Zhaoel2023,Ren2019,Tang2016}. 
%added 	three citations Zhaoel2023,Ren2019,Tang2016
	
%	\color{blue}
	The scattering of electrons by gate-defined QDs with sharp p-n junction has been theoretically studied using the Mie scattering \cite{Heinisch2013} and multiple scattering theory \cite{Zhaoel2023,Ren2019,Tang2016}. In comparison to those approaches, in this paper using first Born approximation, we show that, the differential scattering cross-section (DSC) of a scattering potential is proportional to the Fourier transform of the potential profile. This is analogous to intensity in the back focal plane in the case of Fraunhofer diffraction pattern \cite{duffieux1983fourier, BornWolf}. However, due to Klein tunneling and absence of back-scattering, the MDFs in graphene are mainly transmitted towards the forward region which makes graphene scattering interesting.  Particularly in the inset of  \figref{figSL} (b) we show the differential scattering cross-section from a single QD forms a cardioid like pattern which combine the effects of Mie scattering and Klein tunneling for the gaussian QD we are considering.  In this framework we showed how using dc-resistivity, an experimentally measurable quantity which is dependent on this DSC, we can analyze the properties of the scattering potential in a way that is reminiscent of optical image processing. While in optical image processing a second lens is used to reconstruct the image (depicted in \figref{figSL}), similar thing is however not possible in our system. To overcome this limitation we have introduced another degree of freedom, rotation angle $\phi$, which is the angle between the direction of propagation of incident plane wave and the symmetry axis of the two dimensional QD lattice (TDQDL). By doing Fourier analysis of this dc resistivity we can get information about the symmetry and lattice configuration, lattice constant, size and location of defect of the scattering lattice. In a moir\'{e} pattern of two TDQDL, we have shown that the Fourier analysis can give us the information about the symmetries (moir\'{e} pattern of two hexagonal/square lattices) and commensurate conditions.
%	\color{black}
	
	\begin{figure*}
		\centering
		\includegraphics[width= 0.8 \linewidth]{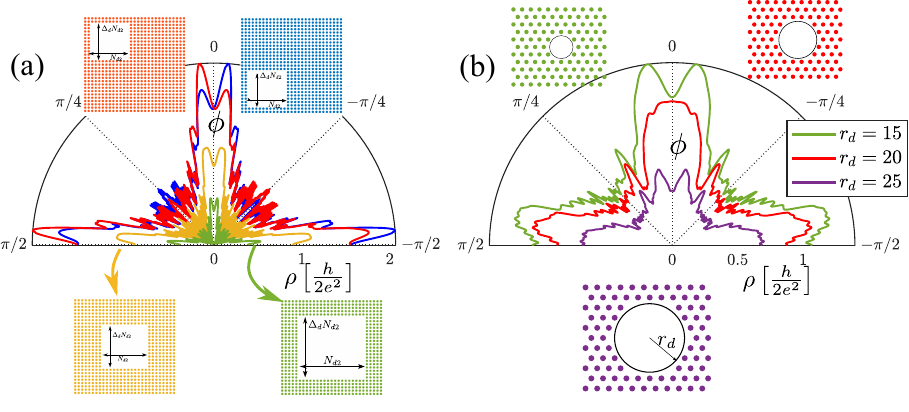}
		\caption{\justifying   Resistivity pattern for (a) square QD lattice of size $N_2=200$ and $\Delta=1$ with square defect region with different sizes and (b) hexagonal QD lattice of size $N_2=61$ and $\Delta=1$ with circular defect region of different radiuses. In (a), The resistivity pattern is symmetric on both sides of $\phi=0$ only when the defect region is centred at the origin and for the blue curve, we have removed scatterers from $n_1=10$ to $110$ and $n_2=10$ to $110$. For the orange curve $n_1=10$ to $110$ and $n_2=90$ to $190$. In (b), the defect region is placed in the centre of the original QD lattice. Here, the resistivity pattern is symmetric on both sides of $\phi=0$.}
		\label{defect}
	\end{figure*}

	Particularly in the ballistic regime, with the Fermi velocity $v_F\sim10^6$ m s$^{-1}$, the mean free path of  the charge carriers in graphene are several microns \cite{novoselov2004electric,neto2009electronic, bhandari2018} $\gg$
	the size of ($\sim$ $10$ nm) \cite{gutierrez2016} such scatterers, and the Fraunhoffer criterion is satisfied \cite{bhandari2018, bhandari2016, berezovsky2010}.
	 %\color{blue}
	In order to observe a clear diffraction pattern in the optical case, the aperture length must be comparable to the wavelength of light \cite{jenkins1976fundamentals, BornWolf}. This is ensured in our analogue solid-state system by considering that the lattice spacing ($\sim 40 $ nm) and characteristic length ($\sim 10$ nm) of each barrier \cite{gutierrez2016} in this QD lattice are smaller than the de Broglie wavelength calculated from the incident energy of the graphene electrons ($\sim 30$~-~$40$ meV) \cite{Forghieri2022}.
	 %\color{black}
	  Also, for the typical energy range we have considered, the wavevector of the MDF is much smaller than the Fermi wave vector,  \ie $k\ll k_F$, eliminating the possibility of inter-valley scattering. Thus, the problem can be modelled by starting with the single-valley non-interacting Hamiltonian of the massless Dirac fermions in single-layer graphene given by  $\hat{H}_0=v_F \mathbf{\sigma} \vdot \mathbf{p}$. 
	Such a two-dimensional QD lattice (TDQDL)  serves as two-dimensional grating for the incident charge carriers for graphene, leading to an electronic analogue of Fraunhofer diffraction pattern, but with an important distinction characterising the absence of backscattering for such MDF \cite{klein,dartora2022theory}.  In \secref{theory}, we discuss the scattering of MDF from three prototype combinations of such dot lattices and lay down the basic premises of our theoretical approach.

%	\color{blue}
	\section{Theory} \label{theory}
%	\color{black}

	The scattered state of such MDF $\ket{\Psi_{\mathbf{k}}}$ from an arbitrary scattering potential $V$ can be obtained by finding out the solutions of $\left(\hat{H}_0+ \hat{V} \right) \ket{\Psi_{\mathbf{k}}}=E \ket{\Psi_{\mathbf{k}}}$ using the Lippmann-Schwinger formalism and can be written as \cite{sakurai, Sadhan}
	
	\begin{equation}
		\Psi_{\mathbf{k}}^{e,+}(\mathbf{r})  = \phi_{\mathbf{k}}^e\left(\vec{r}\right) - \frac{e^{ikr}}{2 \hbar v_F} \sqrt{\frac{ik}{\pi r}} \bmat
		1 \\
		e^{i \theta}
		\ebmat\mel**{\phi_{\mathbf{k'}}^e}{\hat{T}}{\phi_{\mathbf{k}}^e} \label{FBA1}
	\end{equation}
	
	where, transition operator $\hat{T}=\hat{V} +\hat{V} \hat{G_0^{\pm}} \hat{V } +\hat{V} \hat{G_0^{\pm}}\hat{V} \hat{G_0^{\pm}} \hat{V} + \ldots$ and $\hat{G}_0^{\pm}$ is the Green's function defined as $\hat{G_0^{\pm}}= \lim_{\epsilon \rightarrow 0} \left[E I_2 - \hat{\mathbf{H}}_0 \pm i \epsilon \right]$. ($\phi_{\mathbf{k}}^{e, h}$) are respectively free particle electron and hole solutions of  $\hat{H}_0 \ket{\phi_{\mathbf{k}}} = E \ket{\phi_{\mathbf{k}}}$, Here, $\mathbf{k}$ is the wave-vector of the incident wave, $\mathbf{k'}=k \hat{r}$ , $\theta$ is the angle between $\mathbf{k}$ and $\mathbf{k'}$ (see \apref{Lippman}).%(see sec. I, SM \cite{supp}). 

		%\endgroup
		%\end{widetext}
	
	Realistic QD lattices that are heavily n-doped centres on a p-doped background \cite{lee2016imaging, Zhao,li2022recent} 
	serves as scattering  potential $V$ as a Gaussian quantum dot array 
	
	\begin{equation}
		V(\mathbf{r})=\sum_n \left(\frac{V_0}{2 \pi \beta^2 }\right) e^{-\frac{1}{2}(\frac{\mathbf{r-r_n}}{\beta})^2} \label{V} 
	\end{equation}
	
\begin{table*}
	\centering
	\caption{Dictionary for various quantities in Fraunhofer diffraction in optics and their counterparts in scattering of MDF in graphene from a QD lattice}
	\label{table}
	\begin{tabular}{l l}
		\hline
		\hline
		\parbox[t]{0.45\linewidth}{\textbf{Fraunhofer diffraction in optics}} &
		\parbox[t]{0.45\linewidth}{\textbf{Scattering of MDF in graphene from a QD lattice}} \\
		\hline
		\hline
		\parbox[t]{0.45\linewidth}{Aperture profile in front focal plane, $A(\mathbf{r'})$}  & \parbox[t]{0.45\linewidth}{Scattering potential $V(\mathbf{r'})$} \\
		\hline
		\parbox[t]{0.45\linewidth}{Field distribution at  back focal plane,  $G(u,v)=\frac{1}{\lambda f}\int \int dx' dy' g(x',y') e^{-i(u x'+v y')}$}  & \parbox[t]{0.45\linewidth}{Scattering amplitude $f(\theta)=- \sqrt{\frac{ik}{2 \pi}} \frac{\left(1+e^{i \theta}\right)}{2 \hbar v_F}  \int \int dx' dy' V(x',y') e^{i(q_x x' + q_y y')}$} \\
		\hline
		\parbox[t]{0.45\linewidth}{Intensity $I=\abs{G(x,y)}^2$}  & \parbox[t]{0.45\linewidth}{Differential Scattering cross section $\derivative{\sigma}{\theta}= \abs{f\left(\theta\right)}^2$} \\
		\hline
		\parbox[t]{0.45\linewidth}{$u=\frac{2 \pi x}{\lambda f}$ and $v=\frac{2 \pi y}{\lambda f}$} & \parbox[t]{0.45\linewidth}{$-q_x$ and $-q_y$}\\
		\hline
		\parbox[t]{0.45\linewidth}{$\left(\frac{2 \pi}{\lambda f}\right) \mathbf{r}$} & \parbox[t]{0.45\linewidth}{$-\mathbf{q}= \mathbf{k'}-\mathbf{k}$} \\
		\hline
		\parbox[t]{0.45\linewidth}{$\left(\frac{2 \pi}{\lambda f}\right)^2 r^2$} & \parbox[t]{0.45\linewidth}{$q^2=4 k^2 \sin[2](\frac{\theta}{2})$} \\
		\hline
		\hline
	\end{tabular}
\end{table*}

%	\begin{table*}
%		\centering
%		\caption{\label{table}Dictionary for various quantities in Fraunhofer diffraction in optics and their counterparts in scattering of MDF in graphene from a QD lattice:}
%		%	\begin{ruledtabular}
%			%		\begin{tabularx}{ P{0.5 \linewidth} P{0.5 \linewidth}  }
%				\begin{tabularx}{\textwidth}{YY}
%					\hline\hline
%					{\bf Fraunhofer Diffraction in optics} &  {\bf Scattering of MDF in graphene from QD lattice }\\
%					\hline
%					\hline
%					\color{black}
%					 Aperture profile in front focal plane, $A(\mathbf{r'})$   & Scattering potential $V(\mathbf{r'})$    \\
%					\hline
%					Field distribution at  back focal plane,  $G(u,v)=\frac{1}{\lambda f}\int \int dx' dy' g(x',y') e^{-i(u x'+v y')}$ &  Scattering amplitude $f(\theta)=- \sqrt{\frac{ik}{2 \pi}} \frac{\left(1+e^{i \theta}\right)}{2 \hbar v_F}  \int \int dx' dy' V(x',y') e^{i(q_x x' + q_y y')}$  \\
%					\hline
%					Intensity $I=\abs{G(x,y)}^2$ & Differential Scattering cross section $\derivative{\sigma}{\theta}= \abs{f\left(\theta\right)}^2$\\
%					\hline
%					$u=\frac{2 \pi x}{\lambda f}$ and $v=\frac{2 \pi y}{\lambda f}$ & $-q_x$ and $-q_y$\\
%					\hline
%					$\frac{1}{\lambda f}$ &  $-\frac{1}{2 \hbar v_F} \sqrt{\frac{ik}{2 \pi}} \left(1+ e^{- i \theta}\right)$\\
%					\hline
%					$\left(\frac{2 \pi}{\lambda f}\right) \mathbf{r}$ & $-\mathbf{q}= \mathbf{k'}-\mathbf{k}$\\
%					\hline
%					$\left(\frac{2 \pi}{\lambda f}\right)^2 r^2$ & $q^2=4 k^2 \sin[2](\frac{\theta}{2})$ \\
%					\hline
%					\hline
%				\end{tabularx}
%			\end{table*}
	
	where $\mathbf{r_n}$s are the centres of each quantum dot where the Gaussian potential of width $\beta$ forms its maxima. Such potential profile can be created using a needle-like electrode offered by an STM tip \cite{on-off,weddingcake,grushevskaya2021electronic,lee2016imaging, Zhao,lee2016imaging,li2022recent},
	connected with a gate potential as shown in \figref{figSL}(a).

	In experimental systems \cite{weddingcake, on-off,lee2016imaging}, these quantum dots are made on graphene/hBN heterostructures on \ch{SiO2}/Si substrate. The gate potential creates a stationary charge distribution in the insulating hBN underlayer, which creates the Gaussian potential profile in the graphene sheet. The \ch{SiO2}/Si substrate acts as a global back gate. To make an array of such QDs, the single electrode can be replaced by an array of such electrodes \cite{wang2022, Schroer2007, hensgens2017, grushevskaya2021electronic, salfi2016}.

	For the type of potential profile depicted in  \eqaref{V}, the premise for Fourier electron optics(FEO) with such MDF can be developed first by evaluating differential scattering cross-sections (DSC), which we did for three carefully chosen  prototype combinations of such dot lattices and subsequently demonstrating their effect on transport. The DSCs evaluated respectively are: 
	\begin{widetext}
		\begin{subequations}\label{eq3}
			\begin{eqnarray}
				\derivative{\sigma}{\theta} &=& 
				\frac{1}{4 \hbar^2 v_F^2} \left(\frac{k}{\pi}\right) V_0^2 \mu(\theta) M_1^2(\theta,\phi) M_2^2(\theta,\phi) \mbox{, square lattice with dimension $(\Delta N_2 \times N_2)$,} \nonumber \\
				& & \hspace{0.35\linewidth}\mbox{and hexagonal lattice rotated in an angle $\phi$} \label{square}\\
				&=& \frac{k V^2 \mu(\theta)}{4 \pi \hbar^2 v_F^2} \left[ M_1 \left(N_2, \Delta,\phi,\mathbf{q} \right) M_2 \left(N_2,  \phi,\mathbf{q}\right) - M_1 \left(N_{d2}, \Delta_{d},\phi,\mathbf{q} \right) M_2 \left(N_{d2}, \phi ,\mathbf{q}\right)\right]^2  \mbox{, square QD lattice with a }, \nonumber \\ 
				&&\hspace{0.45\linewidth}\mbox{defect region of rectangular shape} \label{squarewithcavity}\\
				&=& \frac{k V^2 \mu(\theta)}{4 \pi \hbar^2 v_F^2} \left[ M_1 \left(N_2, \Delta,\phi - \frac{\delta}{2},\mathbf{q} \right) M_2 \left(N_2,  \phi- \frac{\delta}{2},\mathbf{q}\right) + M_1 \left(N_2, \Delta,\phi + \frac{\delta}{2},\mathbf{q} \right) M_2 \left(N_2, \phi + \frac{\delta}{2},\mathbf{q}\right)\right]^2 \nonumber \\
				&&\hspace{0.4\linewidth} \mbox{  moir\'{e} pattern of two QD lattices }\label{squaremoire}
			\end{eqnarray}
		\end{subequations}
	\end{widetext}
	
	% \clearpage
	%\begin{widetext}
	%\begingroup
	%\renewcommand{\arraystretch}{1} % Default value: 1

			%\begin{figure}%[H]
			%	\centering
			%	\includegraphics[width=0.75\columnwidth]{diffscat_1.pdf}
			%	\caption{\justifying \color{red}DSC for a single QD is plotted as a function of $\theta$ for different values of width of gaussian potential ($\beta$).  }
			%	\label{diffscat}
			%\end{figure}
			
			\begin{figure*}
				\centering
				\includegraphics[width= \linewidth]{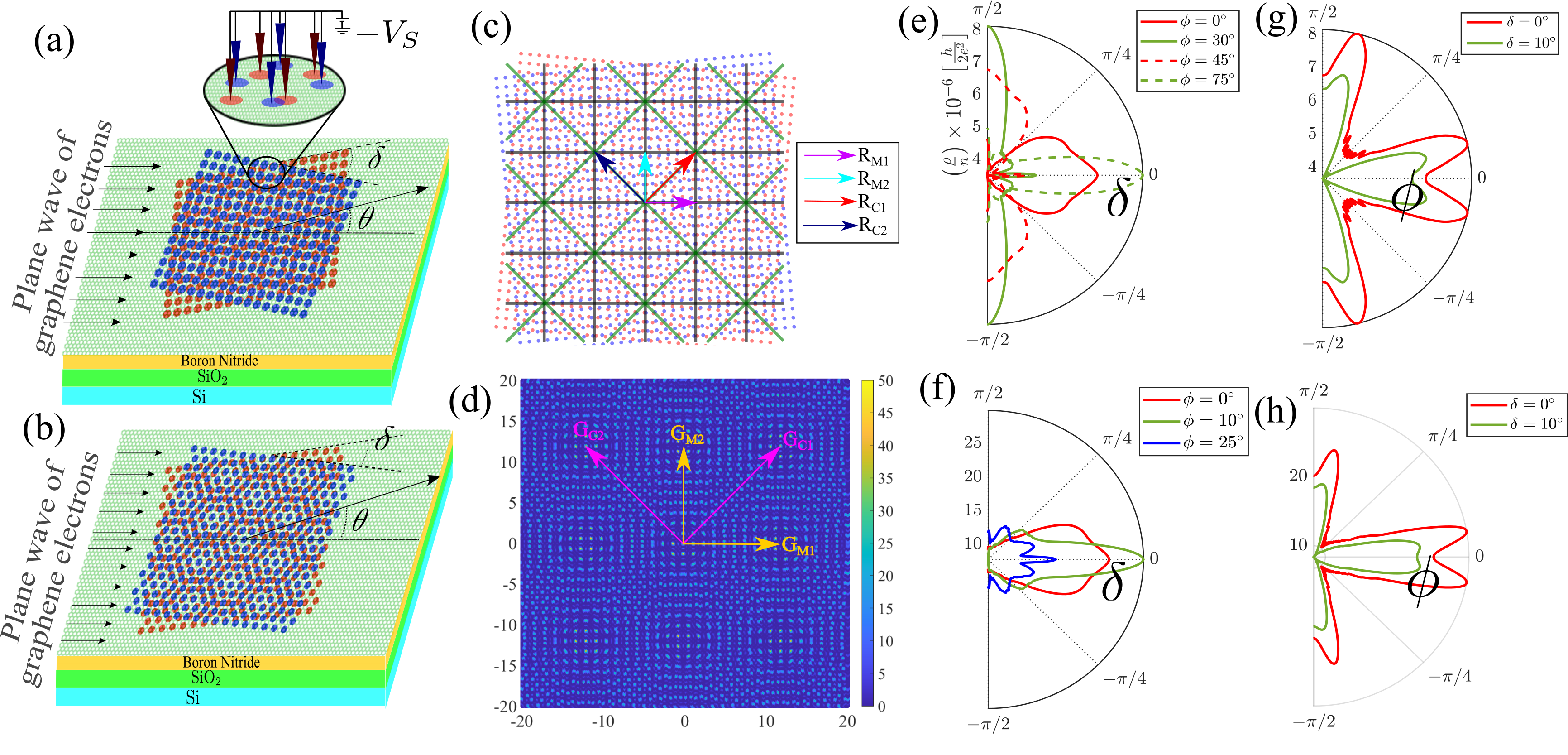}
				\caption{\justifying  The schematic diagram of a plane wave of massless Dirac fermions getting scattered by a moir\'{e} superlattice of two (a) square and (b) hexagonal lattices of Gaussian quantum dots(QD) in graphene. Moir\'{e} pattern made by two square lattices of TDQDL producing a commensurate structure at a twist angle, $\delta \approx 6.026^{\circ}$ is shown in (c).  The moir\'{e} lattice is shown in green, and the commensurate lattice is shown in black. Such quantum dots can be created in experimental system by using tips with applied gate voltage in the same way as in \figref{figSL}. In (d) $\abs{\Tilde{V}(\mathbf{q_1})}^2$ is plotted as a function of $q_{1x}$ and $q_{1y}$ for the above scattering potential.  The resistivity pattern with fixed the mean angle($\phi$) is shown in (e) and (f) for a TDQDL scattering potential made with moir\'{e} pattern of two square and hexagonal lattices, respectively. In (g) and (h), the resistivity pattern is plotted with fixed twist angle($\delta$) again for a moir\'{e} pattern of two square and hexagonal lattices, respectively.}
				\label{figM}
			\end{figure*}
		
			$M_1(N_2, \Delta,\phi,\mathbf{q})=\sin(N_2 \Delta d  \mathbf{q}\vdot \hat{x'}/2)/\sin(d \mathbf{q}\vdot \hat{x'}/2)$ and  $M_2(N_2,\phi,\mathbf{q})= \sin(N_2 d \mathbf{q}\vdot \hat{y'}/2)/\sin(d \mathbf{q}\vdot \hat{y'}/2)$ in \eqaref{eq3} are conventional Fraunhofer diffraction patterns for an one-dimensional grating in mutually transverse direction. $\mu(\theta)=(1+ \cos\theta) \exp[-4 k^2 \beta^2 \sin[2](\theta/2)]$ is due to the gaussian aperture profile (\eqaref{V}) modulated by a factor due to the absence of backscattering \cite{kat,klein} for MDFs in graphene , and $\mathbf{q}=\mathbf{k}-\mathbf{k'}$, $\mathbf{r}_n=n_1 d \hat{x'}+n_2 d \hat{y'}$, where $n_1$ and $n_2$ are integers, and the $x', y'$ axis is rotated with respect to the incident electrons by an angle $\phi$.
			For the first case, we take QDs arranged in square and hexagonal lattice arrangement with suitable choices of $\mathbf{r_n}$.
			The corresponding result, which is plotted in \figref{figSL}(b) in a polar plot, mimics the well-known Fraunhofer pattern with suitable modification due to the absence of back-scattering. 
			This can be thought of as a two-dimensional generalisation of the well-known result of scattering by a smooth p-n junction in one dimension \cite{Cheianov2006}, $T \sim \exp(-\pi k_F \beta \sin[2](\theta)/2)$, as the scattering decreases exponentially with $\beta$ in \eqaref{square} as $\derivative{\sigma}{\theta} \sim e^{-4 k^2 \beta^2 \sin[2](\theta/2)}$ with a characteristic length($\beta$ in our case).
%			\color{blue}
			In the inset of \figref{figSL} (b), we show a polar plot of DSC from a single quantum dot. Due to Klein tunnelling, we can observe that the DSC is maximum in the forward direction and zero in the backward direction due to absence of backscattering \cite{kat,klein}.
			% In sec. II of SM \cite{supp},
			In \apref{2des}, we compare DSC of electrons scattered by TDQDL in non-relativistic two-dimensional electron systems. In that case the absence of backscattering like the MDFs in \eqaref{eq3} is absent. In case of MDFs in graphene the scattered MDFs are mostly confined in the forward direction \cite{kat,klein}.  The comparison with the corresponding cases of non-relativistic charge carriers is discussed in  \apref{2des}.
			
			% \cite{supp}.

			In the next two cases, we first consider a scattering lattice made of TDQDL with defects of different shapes and sizes (\eqaref{squarewithcavity}, \figref{defect} ), and then a third case of scattering potential made with moir\'{e} pattern of two square lattices of TDQDL (\figref{figM}), making the differential scattering cross-section dependent on the twist angle($\delta$) between the two square TDQDLs.
			The scattering amplitude for the TDQDL with a defect region  is the difference between the contribution from the original TDQDL scattering potential (without the defect region ) and a TDQDL of the same shape and position of the defect region in \eqaref{squarewithcavity}. In the third case of moir\'{e} pattern, the scattering amplitude is again the sum of the contributions from two lattices rotated with respect to each other, and the differential scattering cross section in \eqaref{squaremoire} is the square of this sum. 
			This happens due to the linearity property of the Fourier transform \cite{BornWolf}. In the analogous case of optical diffraction systems, this is known as Babinet's principle  \cite{babinet1837, BornWolf, Jimenez2001}.
			
			For the square TDQDL with defect region, we have considered four specific cases by varying the locations and sizes of the cavities. For a square defect region ($(\Delta_{ d }=1)$) in a square TDQDL, the length of the side of the square-shaped cavities($N_{ d2 }$) is varied to vary the size of the cavities. We have also taken two cases where the centre of the defect region does not coincide with the centre of the parent square TDQDL.
			For the moir\'{e} pattern, the moire lattice vector $d_{M}$ is related to the twist angle by $\delta= 2 \sin[-1](\frac{d}{2 d_M})$ \cite{Dunbrack2023}.
			The Bragg condition for the first two cases in \eqaref{square} and \eqref{squarewithcavity} is given by $k^2 d^2 \sin[2](\frac{\theta}{2})= (n_1^2 + n_2^2) \pi^2 $, where $n_1$ and $n_2$ are integers (see \figref{sing} in \apref{max}). %\color{black}
			%(see Fig. S2 in sec. III of SM \cite{supp}).

			The moir\'{e} pattern produces a periodic pattern with lattice periodicities $\sqrt{2N} d_M$ \cite{Dunbrack2023, Can2021}for the commensurate angles where $N$ is an integer. 
			The commensurate super-cell lattice vectors \cite{Santos2007, Carr2020}
			for $\phi=0$ are given by $\mathbf{R}_{C1}=\left(\frac{d}{2 \sin(\delta/2)}\right) N(\hat{x}+\hat{y})$ and $\mathbf{R}_{C2}=\left(\frac{d}{2 \sin(\delta/2)}\right) N(-\hat{x}+\hat{y})$ 
			 
			For every value of N, we get a set of commensurate angles. For the particular commensurate structure that we have considered for square lattice, this common period is larger than the moir\'{e} cell by factor $\sqrt{2}$ \cite{Dunbrack2023} ( see \figref{figM}(c)). The moir\'{e} lattice with lattice periodicity $d_M$ is shown in green. This commensurate lattice is shown in black in \figref{figM}(c). The primitive lattice vectors $\mathbf{R}_{C1}$ and $\mathbf{R}_{C2}$ are not same as the moir\'{e} lattice vectors $\mathbf{R}_{M1}=\left(\frac{d}{2 \sin(\delta/2)}\right) \hat{x}$ and $\mathbf{R}_{M2}=\left(\frac{d}{2 \sin(\delta/2)}\right) \hat{y}$. Due to the commensurate periodicity, scattering potentials made with such patterns show additional Bragg condition from \eqaref{squaremoire},
			\begin{equation}\label{mbragg}
				2 N k^2 d_M^2 \sin[2](\frac{\theta}{2})= (m_1^2+m_2^2)
			\end{equation}
			 
			The detailed discussion on the maximum scattering conditions is shown in \apref{max}. %\color{black} %sec. III of SM \cite{supp}.

			\begin{figure*}
				\centering
				\includegraphics[width=1\linewidth]{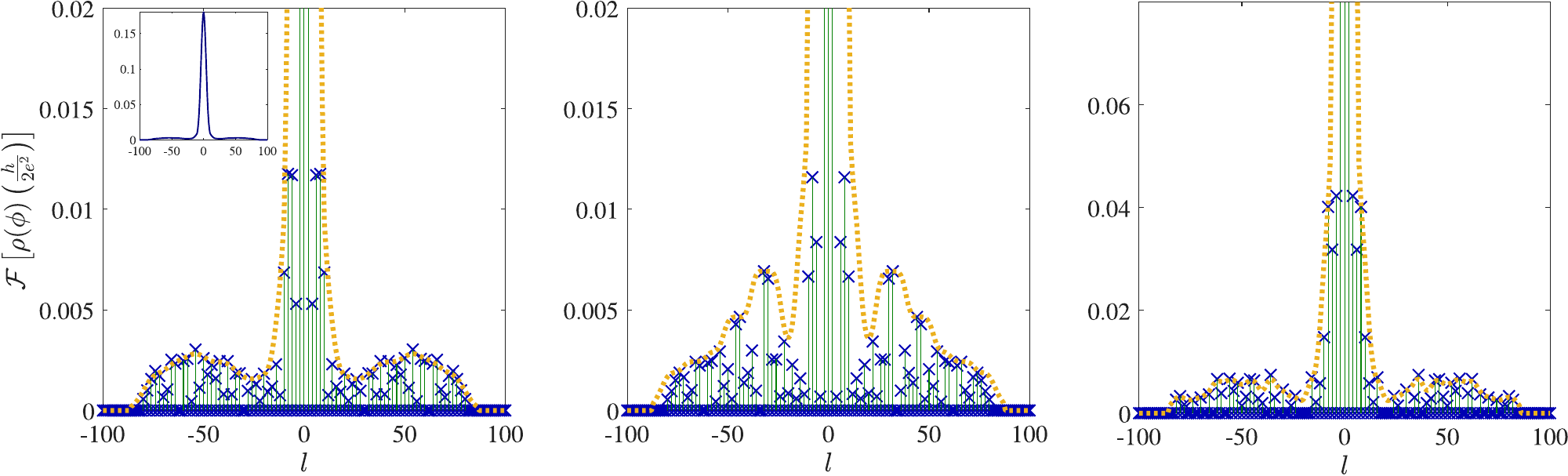}
				\caption{\justifying (a) Shows the Fourier transform(FT) of the resistivity pattern for a TDQDL with $N_{2}=50$, $\Delta=1$ and $d=70$(nm). The blue cross ($\times$) denotes the value of amplitude corresponding to each spatial frequency ($l$). In the inset, we have shown the total data. The main figures do not show the central peak to display the smaller values. The FT of the resistivity pattern through a Gaussian filter for the same TDQDL scattering potential with a square defect region (in the centre) is shown in (b). In (c), we show the FT of the resistivity pattern through a Gaussian filter for a scattering potential made with a moir\'{e} pattern of two square TDQDL with the same lattice constant.}
				\label{ft}
			\end{figure*}

			To understand how the above results can be interpreted as an electronics analogue of Fourier optics in a succinct way, in the left column of \tref{table} we list quantities that characterise a Fraunhofer diffraction-based spatial frequency filtering in optical imaging systems \cite{Birch1968, KGBirch_1972, goodman2005introduction} and are demonstrated in Figs.~\ref{figSL} (e), (f), (g). In the right column of the same table, we list the corresponding quantities that define the analogue system of MDF scattered by a QD lattice.
			From \tref{table}, we can observe that both the scattering amplitude in our system, $f(\theta)$ and the amplitude distribution at the spatial frequency plane in \figref{figSL}(e), $G(u,v)$ is the Fourier transform of the scattering potential $V(x', y')$ and the object distribution $g(x', y')$ respectively, whereas the presence of $(1+e^{i \theta })$ term in $f(\theta)$ of MDF indicates the absence of backscattering \cite{kat, klein, Stander2009}. For MDF, the position coordinate ($\mathbf{r}$) is transformed to angular wave number coordinate $(\mathbf{q})$, whereas in the corresponding optical system,  the coordinates of the plane of aperture, $x'$ and $y'$, are transformed to $u$ and $v$, respectively. Correspondingly, the differential cross section, $\derivative{\sigma}{\theta}$ in the case of MDF is replaced by intensity distribution $I$ in the optical case as demonstrated in Figs.~\ref{figSL}(b) and (f),

			A noteworthy difference with the corresponding optical system appears when the superposition of two such TDQDL, rotated with respect to each other so that the resulting pattern is a moir\'{e} pattern, widely studied in optics \cite{bracewell1995, Ezra1985, Ushkov2020, Liu2017} and condensed matter \cite{Bistritzer2011,Carr2020,Santos2007,Kim2017,Yoo2019,Tarnopolsky2019,Aggarwal2023}.   Such moir\'{e} patterns are studied in optical imaging to improve the microscope's imaging capabilities in Structural Imaging Microscopy \cite{Saxena2015, Strohl2016, Classen2017}. In optical systems, the resultant aperture profile for two transparencies $g_1(x', y')$ and $g_2(x', y')$ is $g_1(x', y')\times g_2(x', y')$ \cite{Zhou2008, Kong2011}. This leads to a new pattern which consists of new beating frequencies \cite{Bryngdahl1974}. However, for MDF scattered by two potentials $V_1(x', y')$ and $V_2(x', y')$, the differential scattering cross-section depends on the Fourier transform of $V_1(x', y')+ V_2(x', y')$. This leads to additional maximum scattering conditions from \eqaref{squaremoire} as shown in \eqaref{mbragg} and depicted in \figref{figM} (d). %\color{black} 
			%However whereas in optical systems the resultant aperture profile becomes $g_1(x', y')\times g_2(x', y')$ \cite{Zhou2008, Kong2011} for two transparencies $g_{1.2}$, making the	aperture plane consists of two transparencies $g_1(x', y')$ and $g_2(x', y')$, leading to the new pattern consists of new beating frequencies \cite{Bryngdahl1974}, for MDF scattered by two potentials $V_1(x', y')$ and $V_2(x', y')$, the differential scattering cross section depends on the Fourier transform of $V_1(x', y')+ V_2(x', y')$ leading to additional maximum scattering conditions from \eqaref{squaremoire} shown in \eqaref{mbragg} and depicted in \figref{figM} (d).
			 The resultant scattering potential shows a new periodicity only for commensurate angles.

			%The above analogy between the scattering problem of MDF from TDQDL and the Fourier optics using a 2D grating naturally leads to the question if this can be extended to the electronic analogue of image processing in the optical case, which is achieved through a second FT through lenses in the optical system ( see  \figref{figSL}(g)). 
%			\color{blue}
			The above analogy between the scattering problem of MDF from TDQDL and the Fourier optics using a 2D grating naturally leads to the question if this can be extended to the electronic analogue of optical image processing. In optical systems this is achieved through a second FT through lenses in the optical system ( see  \figref{figSL}(g)).	In our system, we introduce another degree of freedom, by rotating our scattering potential about a transverse axis passing through the centre of the TDQDL. We define a spatial angle $\phi$ between the propagation direction of the MDF and the symmetry axis of the TDQDL in the graphene plane. While a direct analogy with the optical image processing is difficult, we show that some information about the structure of the TDQDL can be retrieved by calculating the dc resistivity at zero temperature for different rotation angles ($\phi$). Using the semiclassical Boltzmann theory, the dc resistivity at zero temperature \cite{Ferreira1, Ferreira2} is given by
%			\color{black}
			%Whereas a direct analogy is difficult, we show here that selected information about the structure of the TDQDL can be retrieved by rotating our scattering potential about a transverse axis passing through the centre of the TDQDL, thereby introducing a spatial angle $\phi$ between the propagation direction of the MDF and the symmetry axis with respect to the incident plane wave to introduce another degree of freedom, and evaluating the dc resistivity at zero temperature is \cite{Ferreira1, Ferreira2} using the semi-classical Boltzmann theory by 
			\begin{equation}\label{42}
				\rho=\frac{2 n v_F  \sigma_{tr}}{e^2 v_F^2 D(E_F)}= \frac{2 n }{e^2 v_F D(E_F)} \int_{0}^{2 \pi}  d\theta \left(1-\cos \theta \right)\derivative{\sigma}{\theta}  .
			\end{equation} 
			We note that 
%			\color{blue}
			\eqaref{42} is valid in the limit of small concentration $\left(n\right)$ of %external potential centres,
%			\color{blue}
			scattering centers due to quantum dot lattice potential.
%			\color{black} 
			Here, $e=$charge of an electron, $v_F=$ Fermi velocity, $D(E_F)=$ density of states at Fermi energy (variation of resistivity with $E_F$ is discussed in \figref{fig5} of \apref{fermi} ) \color{black} and  $\sigma_{tr}$ is the transport scattering cross-section. Corresponding values of the resistivities in systems considered are within the typical range of observed resistivities in graphene-based systems \cite{chen2008, Balci2012, giubileo2017}. Additionally, our calculated values of resistivities can also be adjusted by choosing the gate voltages applied at every QD and varying the height of the scatterers.

			The corresponding results, namely the angular distribution of resistivity, are plotted in Figs.~\ref{figSL}(c) and (d) for square and hexagonal TDQDL, in Figs.~\ref{defect}(a) and (b)  for TDQDL with a defect region , and in  Figs.~\ref{figM} (e),(f),(g) and (h) for moir\'{e} pattern of two TDQDL lattices, to demonstrate how certain structural information of these TDQDLs can be retrieved from this angle-resolved resistivity. 
			
			\begin{figure}%[H]
				\centering
				\includegraphics[width= 1 \columnwidth]{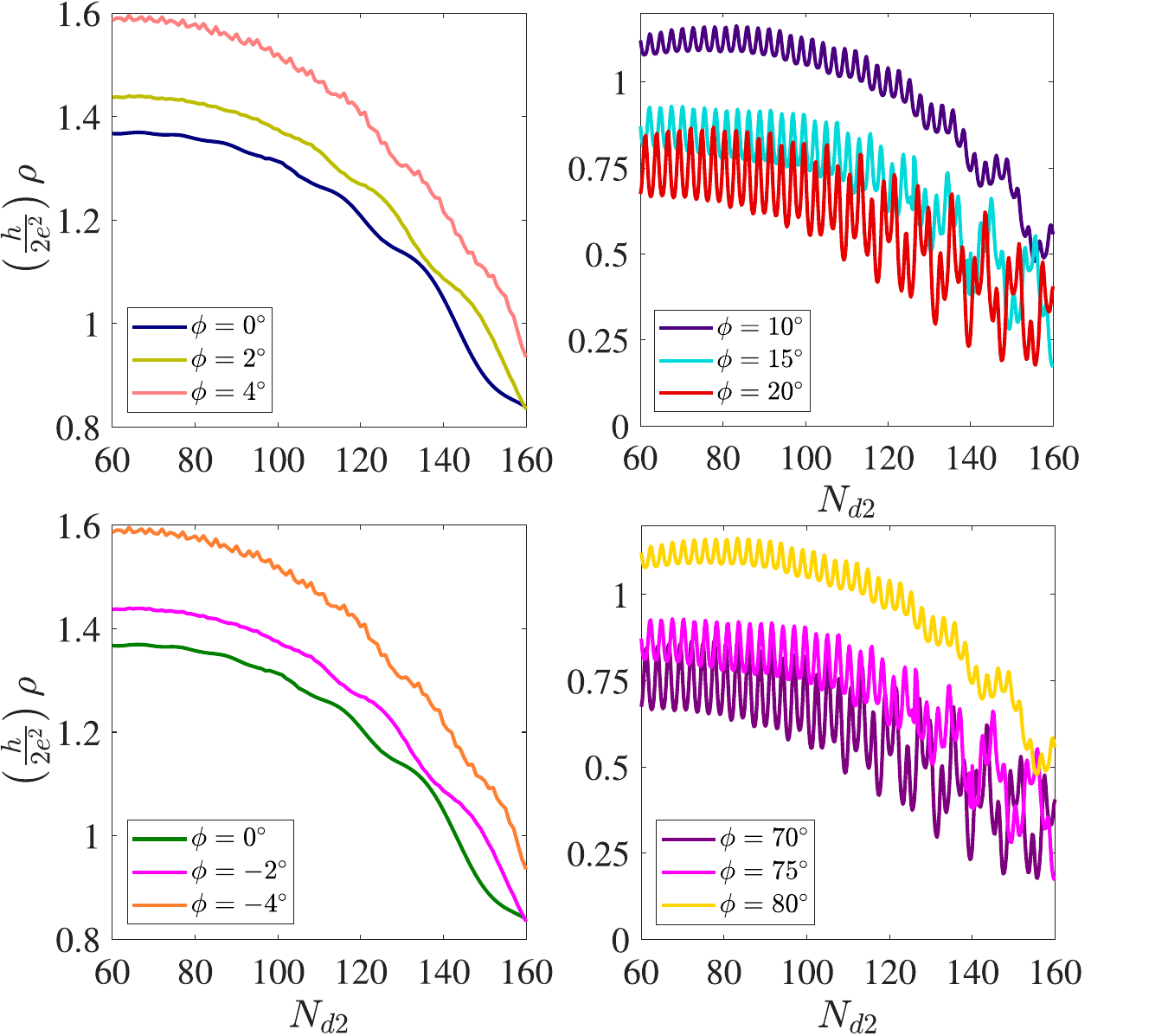}
				\caption{\justifying Resistivity  vs $N_{d2}$ plot for different values of $\phi$. }
				\label{cav}
			\end{figure}

			We begin with the observation that $\rho(\phi)$ in Figs.~\ref{figSL}(c) and (d) reflects the discrete rotational symmetry of the square and hexagonal lattice, e.g. $\derivative{\sigma}{\theta}\at[\Big]{\Delta, N_2,\phi=0\degr}= \derivative{\sigma}{\theta}\at[\Big]{\frac{1}{\Delta},\Delta N_2,\phi=90\degr}$ in case for square lattice, but not for the hexagonal lattice.
			This may be contrasted with the observation in Figs.~\ref{figSL}(c) and (d) that $\rho(-\pi/2)=\rho(\pi/2)$ for both square and hexagonal lattice of TDQDL. For TDQDL with a defect region , the resistivity pattern calculated with the help of  \eqaref{squarewithcavity} and 
			\eqaref{42} reveals both the symmetry of the TDQDL, as well as the location and size of the defect region . In \figref{defect}(a) comparing between blue, red, and orange-yellow plots of the $\rho(\phi)$, we see that the location of the square defect region in the corresponding square TDQDL indeed exhibits itself in the symmetry of the resistivity plot about the $\phi=0$ line. Similarly, comparing the cases of orange-yellow and the green plots of $\rho(\phi)$, we see that the size of the defect region , which is proportional to the number of removed scattering centres, indeed affects the magnitude of the resistivity pattern. Hexagonal TDQDL, with cavities of circular shape with radius($r_d)$ in \figref{defect}(b), shows the same effect
			%(see Fig. S4 and S5 in sec. V of SM \cite{supp}).
			(see \apref{cav_app}). In \figref{cav}, we show the dependence on resistivity at different values of $\phi$ on the size of the defect region. From \figref{cav}, we can observe that the plot is the same for angles $\phi$, $(\pi/2-\phi)$ and $-\phi$. As both the original scattering lattice is made of square lattice and of square shape$(\Delta=1)$ and the defect region is also in the shape of a square, this symmetry comes.  
			
			The differential scattering cross section for the scattering potential made with a moir\'{e} pattern of two square lattices of TDQDL is shown in \eqaref{squaremoire}. In \figref{figM} (d), we show $\Tilde{V}(\mathbf{q})$ 
			as a function of $q_{1x}$ and $q_{1y}$ (see \apref{max}). As expected, FT is also a moir\'{e} pattern of two square patterns in inverse space but multiplied by the same  $\mu(\theta)$ given in \eqaref{squaremoire}. In \figref{figM}, (e), we show the dependence of the resistivity pattern on the twist angle($\delta$) for fixed values of average angle ($\phi$) with the incident plane wave of massless Dirac fermions.

			\begin{figure}%[H]
				\centering
				\includegraphics[width=  \columnwidth]{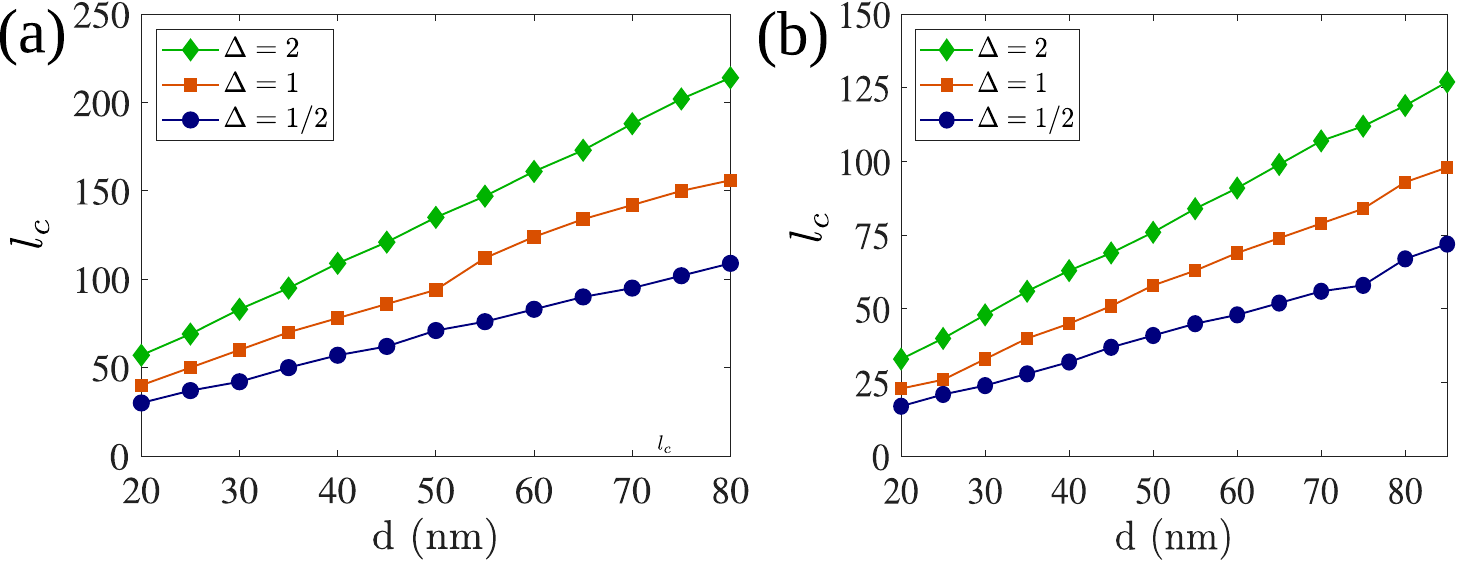}
				\caption{\justifying Cut-off frequency( $l_C$ ) vs lattice constant($d$) plot in the case for (a) square and TDQDL with $N_2=100$(b) hexagonal TDQDL with $N_2=30$. }
				\label{cut-off}
			\end{figure}

			The most prominent features of these resistivity plots of TDQDL in Figs.~\ref{figSL}(c) and (d), for TDQDL with cavities in Figs.~\ref{defect}(a), and (b), and for moir\'{e} pattern of TDQDLs in Figs.~\ref{figM}(e)-(h), are oscillations in resistivity as a function of the angular variable (see \apref{taylor}). 
			To understand the oscillations (see \apref{F}) in resistivity plots Figs.~\ref{figSL}(c) and (d) in a more quantitative way we did an FT of the resistivity that can provide the range of these angular frequencies (see \figref{ft}). If we choose  $l$ as the conjugate variable of $\phi$, we can denote the highest angular frequency component present in one $\rho$ vs $\phi$ curve as $l_c$ or cut-off angular frequency for a given lattice.% \color{black}
			
			%(see Fig. S7 in sec. VII of SM \cite{supp}). was removed

			 For the single layer of TDQDL, $l_c$ depends on the lattice parameter of the scattering lattice as the DSC also depends on the lattice parameter in \eqaref{square}. In \figref{cut-off}, we show the dependence of cut-off frequency on the lattice constant both for square and hexagonal lattice. Here, we see that at a particular value of $d$,  the cut-off frequency is different for different $\Delta$ values. 
			 %As discussed earlier, this happens as $l_P$ changes with $\Delta$. 
			 In \figref{ft} (a), the Fourier spectrum of the resistivity pattern as a function of $l$ 
			for a TDQDL shows a feature that is similar to the side-band formation. We highlight this aspect through an envelope function ( dotted orange curve) over the actual result.
			For the same TDQDL, with a defect region at the centre, and for the same cut-off frequency, this side-band like feature becomes more prominent in \figref{ft} (b), whereas this side-band like feature gets highly suppressed in the case of moir\'{e} pattern of two relatively twisted TDQDls in \figref{ft} (c). These points out that certain angular frequency components get enhanced or suppressed as the scattering region is removed or added, a phenomenon akin to the spatial frequency filtering in Fourier optics, but now happening in the solid state environment for the angle-resolved resistivity, a transport coefficient. This can also be linked to the electronic analogue of Babinet's principle that we reported earlier in this manuscript.
%			\color{blue}
			In the ballistic transport regime, the spatial frequency filtering in the angle resolved resistivity of MDFs scattered from TDQDL is shown as an electronic analogue of Babinet's principle. This forms one of the most prominent findings in this work.
%			\color{black}
			 %Establishing the electronic analogue of Babinent's principle and the consequent spatial frequency filtering in the evaluated angle resolved resistivity in the ballistic transport regime of MDF scattered from TDQDL, thus forming the most prominent findings in this work. 
			
%			\color{blue}
			\section{Discussion}
%			\color{black}

			To summarize, using the Lippmann-Schwinger formalism, we established an electronic analogue of Fourier optics \ie  FEO by mapping the scattering cross-section of MDF from TDQDLs to the Fraunhauffer diffraction pattern and providing a dictionary of such mapping. By considering TDQDL with defect region and moir\'{e} pattern of such TDQDLs, we demonstrated an electronic analogue of Babinet's principle. Harnessing this analogy further with an eye on practical application, we evaluated the angle-resolved resistivity of these scattered MDs, and Fourier analyze the same to show that the Fourier spectrum shows spatial frequency filtering consistent with Babinet's principle.
			With the quantum dot arrays now routinely produced in semiconductor-based systems \cite{wang2022, Schroer2007, hensgens2017}, we hope our proposed analogy can be used for making new electronic devices based on such analogue FEO.
			
%			\color{blue}
			\section{Acknowledgment}
%			\color{black}
			
			We thank Kedar Khare, Rohit Narula and Deepanshu Aggarwal for the helpful discussions. SG and RM are supported by a SPARC Phase II (MHRD, GOI, Project Code P2117)  grant, whereas PSB is supported by a MHRD Fellowship. 
			
%			\color{blue}
			\section{Data Availability Statement}
%			\color{black}
			
			All data that support the findings of this work is included in the main manuscript and supplementary file.
			
%			\color{blue}
			\section{Conflict of Interest}
%			\color{black}
			
			There is also no competing interest between the authors.

		\appendix
		
%		\color{blue}
		\section{Lippmann-Schwinger Integral Equation and Transition Operator} \label{Lippman}
		
		In the presence of a scattering potential $V(\mathbf{r})$, the scattered wavefunction for MDF can be written using the Lippmann-Schwinger formalism \cite{sakurai, Sadhan} as,
		
		\begin{equation} %\label{12}
			\Psi_{\mathbf{k}}^{\eta,\pm}(\mathbf{r})=\phi_{\mathbf{k}}^\eta(\mathbf{r})+\int d^2 \mathbf{r'} G_0^{\pm} (\mathbf{r},\mathbf{r'},E) V(\mathbf{r'}) \Psi_{\mathbf{k}}^{\eta,\pm}(\mathbf{r'})
		\end{equation}
		
		The Green's function for the physically relevant outgoing wave case for MDF is given by,
		
		\begin{equation}
			G_0^+ (\mathbf{r},\mathbf{r'},E) =   - \frac{1}{\hbar v_F} \sqrt{\frac{ik}{8\pi r}} e^{ikr} e^{-i \mathbf{k'} \vdot \mathbf{r'}} \begin{bmatrix}
				1 & e^{-i \theta} \\
				e^{i \theta} & 1
			\end{bmatrix}
		\end{equation}
		
		In the case of First Born Approximation(FBA), the transition operator is approximated till the first term. Which makes the scattering amplitude 
		
		\begin{equation}
			f^{(1)}= - \frac{1}{2 \hbar v_F} \squaret{\frac{i k}{2 \pi}} \left(1+ e^{-i \theta}\right) \tilde{V}(\mathbf{k}-\mathbf{k'}).
		\end{equation}
		
		The asymptotic scattered wave function is given by 
		
		\begin{equation}
			\Psi_{\mathbf{k}}^{e,+}(\mathbf{r})  = \phi_{\vec{k}}^e\left(\vec{r}\right) + \frac{1}{\sqrt{2}} \begin{bmatrix}
				1 \\
				e^{i \theta}
			\end{bmatrix} \frac{e^{ikr}}{\sqrt{r}} f(\theta) 
		\end{equation}
		
		\section{Differernce between Scattering of 2DES and Graphene Electrons} \label{2des}

		The differential scattering cross sections for electrons in a two-dimensional electron system(2DES) scattered from a square TDQDL is given by 
		
		\begin{equation}
			\derivative{\sigma}{\theta} = \left( \frac{2 m}{\hbar^2 }\right)^2 \left( \frac{V_0^2}{8 \pi k}\right) e^{- 4 k^2 \beta^2 \sin[2](\frac{\theta}{2})} M_1^2(\theta,\phi) M_2^2(\theta,\phi). \label{IV001}
		\end{equation}
		
		\begin{figure}%[H]
			\centering
			\includegraphics[width=0.8 \columnwidth]{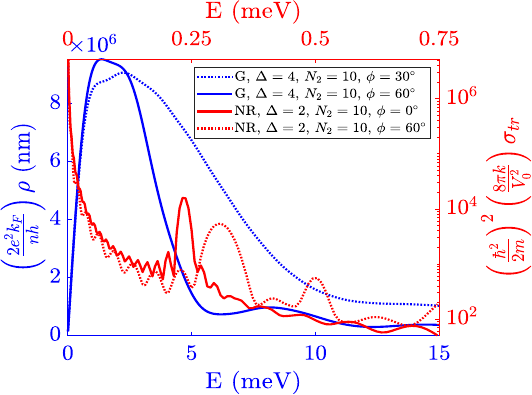}
			\caption{\justifying The energy dependence of the resistivity for graphene(G) and non-relativistic(NR) two dimensional systems.}
			\label{fig3}
		\end{figure}
		
		By comparing \eqaref{square} (main text) and \eqref{IV001}, we can observe that the differential scattering cross section in graphene is proportional to $k(\sim E)$, but in 2DES, it is proportional to $1/k $ $(\sim E^{-1/2} )$~\cite{P_G_Averbuch_1986, Zoubi2005, Ngampruetikorn}. This difference of energy dependence can be understood from \figref{fig3}, where we show the transport scattering cross section for massless Dirac fermions both in the case of graphene and two-dimensional non-relativistic systems.  We can observe from \eqaref{square} that for scattering at low energy, the differential scattering cross section for massless-Dirac fermions becomes $\derivative{\sigma}{\theta}=\frac{k \Delta^2 N_2^4 V_0^2}{4 \pi \hbar^2 v_F^2}\left(1+\cos \theta\right)$. That makes the transport scattering cross section $\sigma_{tr}=\left(\frac{\Delta^2 N_2^4 V_0^2}{4  \hbar^2 v_F^2}\right) k$ and the resistivity becomes, $\rho= \left(\frac{n \pi^2 \Delta^2 N_2^4 V_0^2 }{2 h e^2 k_F v_F^2 } \right) k$. Due to this, we see the linear variation of resistivity with $E$ at low energies in \figref{fig3}.  From \eqaref{IV001} for non-relativistic electrons in 2D, at low energy the differential scattering cross section becomes, $\derivative{\sigma}{\theta}=\left(\frac{2 m}{\hbar^2}\right)^2 \left(\frac{V_0^2}{8 \pi k }\right) \Delta^2 N_2^4$. The transport scattering cross section becomes $\sigma_{tr}=\left( \frac{m V_0 \Delta N_2^2 }{\hbar^2} \right)^2 \times \frac{1}{k}$. In \figref{fig3}, we see that at low energies, the transport scattering cross section becomes inversely proportional to the energy of the incident particle.

		\section{Condition for Maximum Scattering in case of moir\'{e} pattern of two TDQDL} \label{max}

		\begin{figure}%[H]
			\centering
			\includegraphics[width=0.9 \columnwidth]{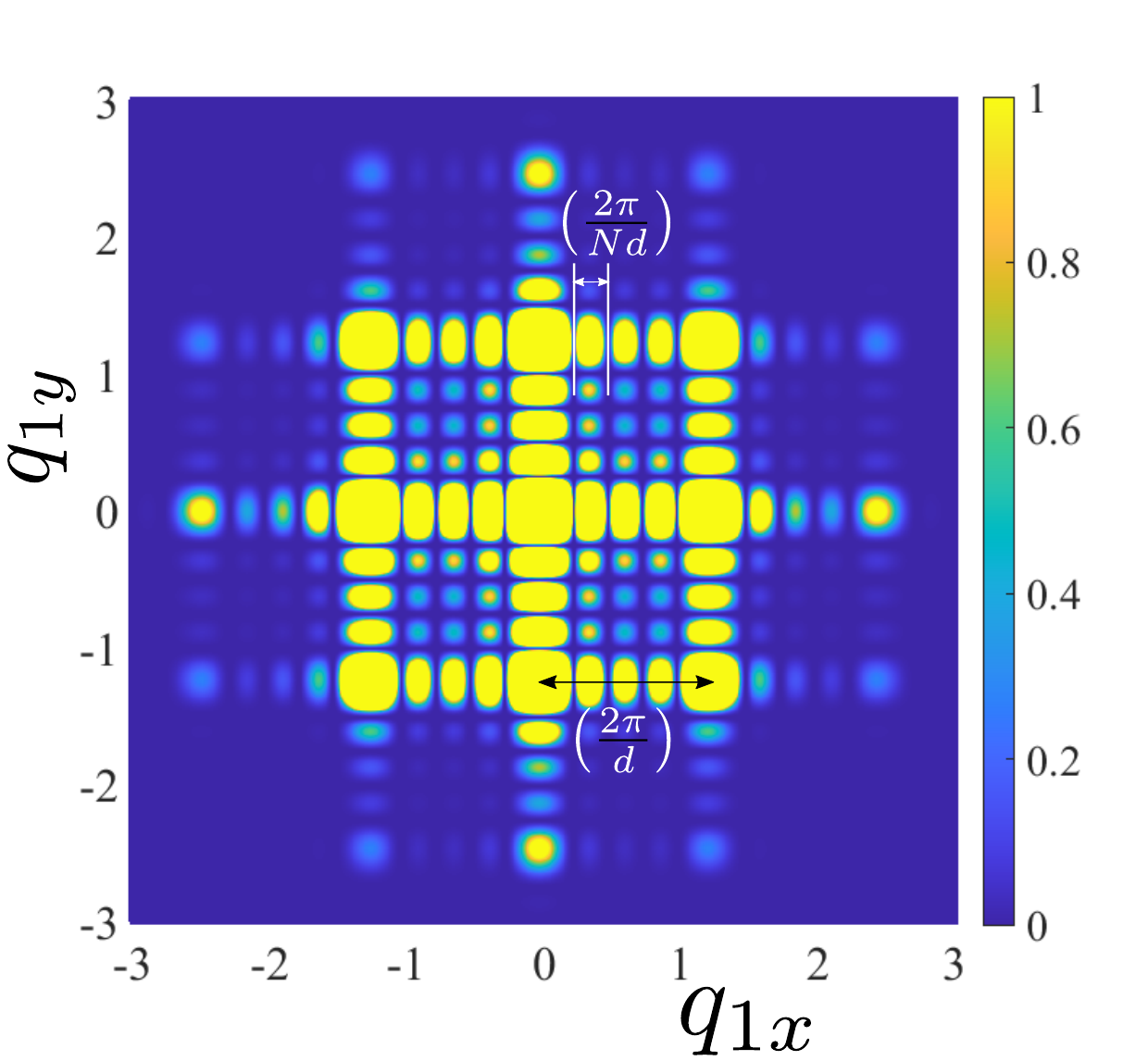}
			\caption{$\abs{\Tilde{V}(\mathbf{q})}^2$ plotted as a function of $q_{1x}$ and $q_{1y}$ for a single layer of TDQDL for $N=5$.}
			\label{sing}
		\end{figure}

		The differential scattering cross-section for these systems is directly proportional to $\abs{\Tilde{V}(\mathbf{q})}^2$. Here $\Tilde{V}(\mathbf{q})$ is the Fourier transform of the scattering potential $V(\mathbf{r})$. 
		In \figref{sing}, we show the Fourier transform of the scattering potential made by a square TDQDL. The positions for the principle maximas are given by $q_{1x}= n_1 (2 \pi /d )$ and $q_{1y}= n_2 (2 \pi /d )$ where $n_1$ and $n_2$ are integers. These correspond to the Bragg condition. The secondary minimas are separated by a distance of $\left(\frac{2 \pi}{N d}\right)$. We can also notice that $\abs{\Tilde{V}(\mathbf{q})}^2$ has maximum value for the central maxima at $q_{1x}=0$ and $q_{1y}=0$ because the Fourier transform of the scattering potential has a Gaussian term $e^{-4 k^2 \beta^2 \sin[2](\theta/2)}$ same as in $\mu(\theta)$ in \eqaref{square} (main text).
		
		For the case of moir\'{e} pattern of two square TDQDL scattering potentials, we have shown the maximum scattering cases in \figref{figM} (d) (main text). We can observe that the Fourier transform is also a moir\'{e} pattern of two square patterns in inverse space. The reciprocal lattice vectors, in this case, is

		\begin{align}
			\mathbf{G_{M1}}  &= \frac{4 \pi \sin(\frac{\delta}{2})}{d} \hat{x}, &  \mathbf{G_{M2}} &= \frac{4 \pi \sin(\frac{\delta}{2})}{d} \hat{y} \\
			\mathbf{G_{C1}}  &= \frac{4 \pi \sin(\frac{\delta}{2})}{d} N(\hat{x}+ \hat{y}), &\mathbf{G_{C2}}&= \frac{4 \pi \sin(\frac{\delta}{2})}{d} N(\hat{y}-\hat{x} )
		\end{align}
		
		Where, $N$ is an integer. For every value of $N$, we have a set of commensurate angles.

		The Fourier transform produces a commensurate pattern only when the scattering potential itself produces a commensurate pattern.	The Bragg conditions, which appear only in the case of commensurate lattice \ie \eqaref{mbragg} correspond to those points in \figref{figM} (d) for which $\mathbf{q}=m_1 \mathbf{G_{C1}}+ m_2 \mathbf{G_{C2}}$.
		
		\begin{figure}%[H]
			\centering
			\includegraphics[width=0.5 \columnwidth]{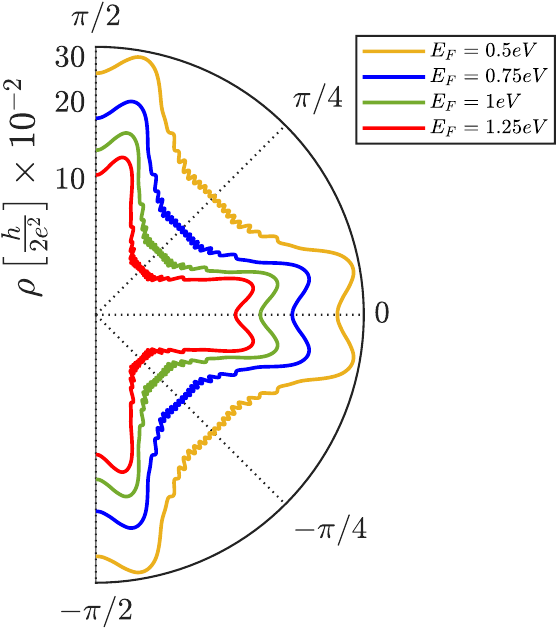}
			\caption{\justifying The resistivity pattern for different values of $E_F$ at $N_2=100$ and $\Delta=1$.}
			\label{fig5}
		\end{figure}
		
		\section{Dependence of Resistivity Pattern on Fermi Energy}\label{fermi}

		For monolayer graphene systems, the fermi energy ($E_F$) and the fermi momentum are controlled by a back-gate voltage $V_g$. If the back gate and the graphene layer are separated by a substrate of  \ch{SiO2} with a width of about $300$nm, the Fermi momentum is given by \cite{Ferreira2},
		
		\begin{figure}%[H]
			\centering
			\includegraphics[width=  \columnwidth]{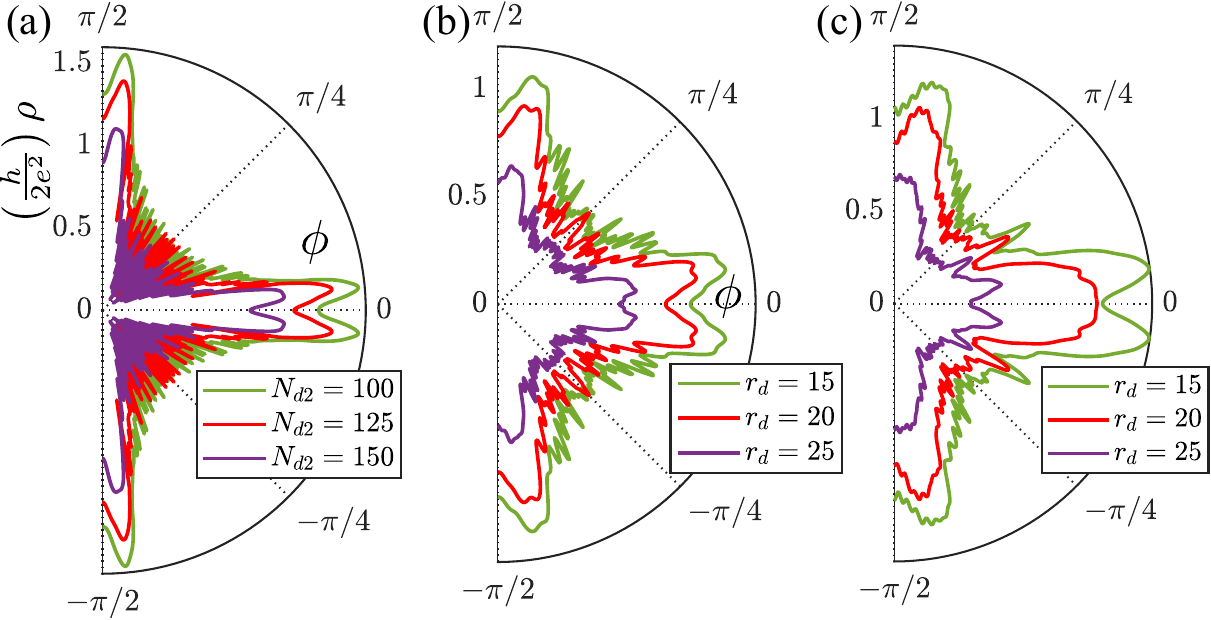}
			\caption{\justifying Resistivity pattern for a TDQDL potential with (a) square defect region in a square TDQDL, (b) circular defect region in a square TDQDL and (c) circular defect region in a hexagonal TDQDL. We have compared defect regions of different sizes.}
			\label{sc}
		\end{figure}
		
		\begin{equation}
			k_F^2=\pi \alpha V_g
		\end{equation}
		
		where, $\alpha= 7.2 \times 10^{-4} $ V$^{-1}$nm$^{-2}$. For a back-gate voltage of $100$ V, the Fermi energy is $0.31$ eV.

		In \figref{fig5},  we show the resistivity pattern for different values of $E_F$ for a fixed configuration. We can observe from here that the value of resistivity decreases with a larger value of Fermi energy \cite{Sohier2014, Efetov2010, chen2008}.

		\section{Dependence of Resistivity Pattern on the Shape of  Defect Region in the Scattering TDQDL Potential.}\label{cav_app}

		Here, we compare the resistivity pattern for TDQDL of different shapes in \figref{sc}. To distinguish the shapes of defect regions, we have to check the symmetries of the original lattice and the defect region.
		
		In \figref{sc}, we can observe that the resistivity pattern depends on the size of the defect region. Now, we can choose a fixed rotation value $\phi$ and compare the resistivity values for different values of $N_{d2}$.

		\begin{figure}%[H]
			\centering
			\includegraphics[width= \columnwidth]{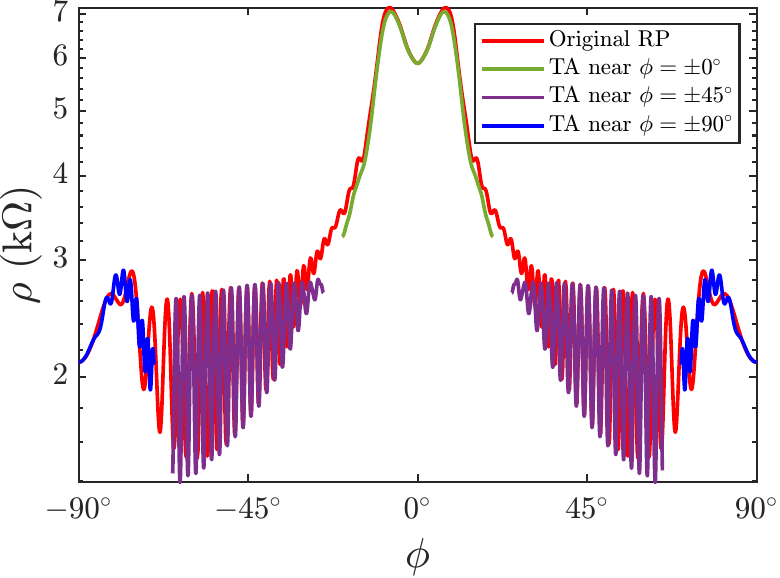}
			\caption{ Resistivity pattern for a TDQDL of dimension $N_2=50$ and $\Delta=2$. The red plot shows the resistivity pattern (RP) calculated from \eqaref{square}, the general $\phi$ dependent differential scattering cross-section. The green, magenta and blue plot shows the approximate value of resistivity near the values of $\phi=0, \pm \frac{\pi}{4}$ and $\pm \frac{\pi}{2}$ using Taylor Approximation(TA). }
			\label{fig6}
		\end{figure}

		\section{Approximations of Resistivity Pattern at Various $\phi$ values for Single Layer of TDQDL}\label{taylor}

		The resistivity plots in Figs.~\ref{figSL}(c) and (d) (main text) as a function of $\phi$ show oscillations having different amplitudes around different angles. To understand them, near $\phi=0, \pm \frac{\pi}{4}, \pm \frac{\pi}{2}$ we evaluate the resistivity with a Taylor expansion around each of these angles, and corresponding approximate plots are then superposed with the exact one evaluated with \eqaref{42} in \figref{fig6}, for a square TDQDL calculated for $N_2=50$ and $\Delta=2$. As expected, they almost overlap near each angle over a range and start showing deviation as one sifts from these angles.

		\section{Oscillations in the Resistivity Pattern} \label{F}

		We observe from Figs.~\ref{figSL} (c) and (d) (main text) that the frequency of oscillations in the $\rho$ vs $\phi$ plot increases with a higher value of $\Delta$ for a fixed $N_2$.  To explain this, note that the only term in \eqaref{square} (main text) that depends on $\Delta$ is $\sin^2\left( N_2 \Delta k d \sin(\frac{\theta}{2}) \sin(\frac{\theta}{2}-\phi)\right)$ . It has a  period(in $\phi$) of oscillation $l_P=\sin^{-1}\left[\frac{\pi}{N_2 \Delta kd \sin(\frac{\theta}{2})}\right]$. Hence, the frequency of oscillation in the resistivity pattern is higher for a higher value of $\Delta$. 
		
		%To understand these oscillations, we have performed Fourier transforms of the $\rho$ vs $\phi$ curve to understand which frequency components are present in the plot. If we choose the conjugate variable of $\phi$ as $l$, then the highest frequency present in the resistivity pattern (cut-off frequency) is $l_c$. In \figref{ft} (main text), we have shown the Fourier transform of the resistivity pattern for the three cases of TDQDL-based scattering potentials we are considering.

		\texttt{Moir\'{e} Pattern of Two QD Scattering Lattices:} 	As the moir\'{e} pattern produced by two square lattices with $\delta= \Delta \delta$ and $\phi=45^{\circ}$ is the same as the moir\'{e} pattern of $\delta=90^{\circ}+ \Delta \delta$ and $\phi=45^{\circ}$, in the resistivity pattern of \figref{figM} (e) we observe that $\rho(\phi,\delta)=\rho(\phi \pm \frac{\pi}{4},\delta \pm \frac{\pi}{2})$. The differential scattering cross section for these two cases is also the same, shown from \eqaref{squaremoire} (main text). 
		Such symmetry is absent in the resistivity of moir\'{e} pattern of two hexagonal TDQDL lattices when plotted as a function of the twist angle ($\delta$) in \figref{figM}(f) (main text).
		Similarly, we observe whereas in \figref{figM}(g) (main text) $\rho(\phi=0^{\circ},\delta)=\rho(\phi=90^{\circ},\delta)$ due to symmetry of the square lattice, this is not the case for the hexagonal counterpart in \figref{figM}(h).
		As the moir\'{e} lattice vector is larger than the lattice vector of the original lattices, the spatial frequencies corresponding to moir\'{e} lattice vector are smaller than the spatial frequency corresponding to the lattice vector of the original lattice. Due to this, the same cut-off frequency analysis which was used for the earlier case in \figref{cut-off} can not detect the dependence of the cut-off on the moir\'{e} lattice vector.

		\color{black}

	\end{document}